\documentclass[twocolumn,showpacs,preprintnumbers,amsmath,amssymb]{revtex4}
\usepackage{txfonts}
\usepackage{mathrsfs}
\usepackage{amssymb}

\usepackage{graphicx}
\usepackage{bm}

\begin{document}

\title{Decoy states for quantum key distribution based on decoherence-free subspaces}

\author{Zhen-Qiang Yin, Yi-Bo Zhao, Zheng-Wei Zhou*, Zheng-Fu Han*, Guang-Can Guo}
\affiliation{Key Laboratory of Quantum Information\\ University of
Science and Technology of China\\ Hefei 230026\\ China}

\begin{abstract}
Quantum key distribution with decoherence-free subspaces has been
proposed to overcome the collective noise to the polarization modes
of photons flying in quantum channel. Prototype of this scheme have
also been achieved with parametric-down conversion source. However,
a novel type of photon-number-splitting attack we proposed in this
paper will make the practical implementations of this scheme
insecure since the parametric-down conversion source may emit
multi-photon pairs occasionally. We propose decoy states method to
make these implementations immune to this attack. And with this
decoy states method, both the security distance and key bit rate
will be increased.
\end{abstract}


\maketitle

\section{introduction}

As a combination of quantum mechanics and conventional cryptography,
Quantum Key Distribution (QKD) \cite{BB84,ekert1991,Gisin}, can help
two distant peers (Alice and Bob) share secret string of bits,
called key. Unlike conventional cryptography whose security is based
on computation complexity, the security of QKD relies on the
fundamental laws of quantum mechanics. Any eavesdropping attempt to
an ideal QKD process will introduce an abnormal high bit error rate
of the key. By comparing subset of the key, Alice and Bob can catch
any eavesdropping attempt. Polarization and phase time of photons
are the most common coding method to implement QKD. But,
birefringence in optical fiber may depolarize the photons, which
makes the polarization coding unsuitable for QKD based on fiber.
Phase time coding is commonly used for fiber QKD. Using "Plug\&Play"
\cite{Gisin} or Faraday-Michelson interferometers\cite{F-M}, phase
time coding can be free from polarization fluctuations due to
birefringence of optical fiber. However, "Plug\&Play" may be
vulnerable for Trojan attack. And for Faraday-Michelson
interferometers \cite{F-M}, it's very sensitive to phase
fluctuations from arms between Alice's and Bob's interferometers. To
overcome this problem, active compensation which makes the system
more complicated and unefficient is used.

 Alternatively, Walton $et$ $al.$ \cite{Walton} proposed a novel QKD protocol based on
decoherence-free space (DFS) and Boileau $et$ $al.$ \cite{Boileau}
developed this scheme to use time-bin and polarization for encoding.
In Boileau's scheme, Alice can encode her qubit in the two-photon
states as follows: $|H\rangle|V\rangle$, $|V\rangle|H\rangle$,
$(|H\rangle|V\rangle+|V\rangle|H\rangle)/\sqrt{2}$, and
$(|H\rangle|V\rangle-|V\rangle|H\rangle/\sqrt{2}$, (in experiment by
J.-W. Pan \cite{DFS experiment}, the four states are:
$(|H\rangle|V\rangle+|V\rangle|H\rangle)/\sqrt{2}$,
$(|H\rangle|V\rangle-|V\rangle|H\rangle)/\sqrt{2}$,
$(|H\rangle|V\rangle+i|V\rangle|H\rangle)/\sqrt{2}$ and
$(|H\rangle|V\rangle-i|V\rangle|H\rangle/\sqrt{2}$), where $H(V)$
means the horizontal (vertical) polarization mode of photons. The
two photons are distinguishable by a fixed time delay $\Delta t_p$,
which is known to Alice and Bob. Then Alice applies a time delay
operation to the $V$ photons and before Bob detects the two photons,
he applies a same time delay operation to the $H$ photons. Finally,
Bob detects the two photons in the $|H\rangle |V\rangle$, $|V\rangle
|H\rangle$ basis or
$\frac{1}{\sqrt{2}}(|H\rangle|V\rangle+|V\rangle|H\rangle)$,
$\frac{1}{\sqrt{2}}(|H\rangle|V\rangle-|V\rangle|H\rangle$ basis.
Due to the fact that
$|\psi^-\rangle=\frac{1}{\sqrt{2}}(|H\rangle|V\rangle-|V\rangle|H\rangle)$
is invariant under collective unitary transformation, this scheme is
insensitive to phase fluctuations from Alice's and Bob's
interferometers. If the interval of the time between the two photons
is just $\Delta t_p$, Bob will successfully get Alice's qubit and
this probability will be $2/3$ assuming the collective noise is
totally random. Besides this, photons from the same pair can provide
precise time references for each other. So in this scheme, accurate
synchronization clock is unnecessary.

 BB84-type QKD protocols which are the most-widely used QKD protocol,
needs single photon source which is not practical for present
technology. Usually, real-file QKD set-ups
\cite{qkd1,qkd2,qkd3,qkd4,F-M} use attenuated laser pulses (weak
coherent states) instead. It means the laser source is equivalent to
a one that emits n-photon state $|n\rangle$ with probability
$P_n=\frac{\mu^n}{n!}e^{-\mu}$, where $\mu$ is average photon number
of the attenuated lased pulses. This photon number Poisson
distribution stems from the coherent state $|\sqrt{\mu}
e^{i\theta}\rangle$ of laser pulse. Therefore, a few multi-photon
events in the laser pulses emitted from Alice open the door of
Photon-Number-Splitting attack (PNS attack) \cite{PNS1,PNS2,PNS3}
which makes the whole QKD process insecure. Fortunately, decoy
states QKD theory \cite{decoy theory1,decoy theory2,decoy
theory3,decoy theory4,decoy theory5}, as a good solution to beat PNS
attack, has been proposed. And some prototypes of decoy state QKD
have been implemented \cite{decoy experiment1,decoy
experiment2,decoy experiment3,decoy experiment4,decoy
experiment5,decoy experiment6,decoy experiment7}. The key point of
decoy states QKD is to calculate the lower bound of counting rate of
single-photon pulses ($S_1^L$) and upper bound of quantum bit error
rate (QBER) of bits generated by single-photon pulses ($e_1^U$).
Many methods to improve performance of decoy states QKD have been
presented, including more decoy states \cite{decoy theory5},
nonorthogonal decoy-state method \cite{nonorthogonal state
protocol}, photon-number-resolving method
\cite{photon-number-resolving method}, herald single photon source
method \cite{herald1,herald2}, modified coherent state source method
\cite{MCS}. And for the intensity fluctuations of the laser pulses,
Ref. \cite{intensity error1} and \cite{intensity error2} give good
solutions.

 As a BB84-type protocol, Boileau's scheme is still vulnerable to
PNS attack. This problem will be discussed in details in the section
II, in which we propose a novel type of PNS attack. In the Section
III, we propose a decoy states method to overcome this problem. In
Section IV, a numerical simulation will be given. Finally, we will
give a summary to end this paper.

\section{PNS attack in Boileau's scheme}

 To implement Boileau's scheme, an ideal two-photon states source which
is far from present technology, is needed. In practice, two-photon
states are generated by parametric down-conversion source(PDCS),
which will emit n-photon ($n>1$) pairs with certain probability.
However, the state from a type-II PDCS can be written like
\cite{PDCS}:
\begin{equation}
\begin{aligned}
|\psiup\rangle=(\cosh\chi)^{-2}\sum_{n=0}^{\infty
}\sqrt{n+1}e^{in\theta}\tanh^n \chi |\Phi_n\rangle,
\end{aligned}
\end{equation}
in which, $|\Phi_n\rangle$ is the state of n-photon pair, given by:
\begin{equation}
\begin{aligned}
|\Phi_n\rangle=\frac{1}{\sqrt{n+1}}\sum_{m=0}^n (-1)^m |n-m,
m\rangle_a |m, n-m\rangle_b
\end{aligned}
\end{equation}
Here, $|n, m\rangle_{a(b)}=|H\rangle^{\otimes n}_{a(b)}
|V\rangle^{\otimes m}_{a(b)}$, a, b means the two spatial output
modes of PDCS respectively. By randomizing the phase $\theta$
\cite{decoy theory1}, we can write the density matrix of the PDCS as
$\rho_\lambda=\int(d\theta/(2\pi))|\psiup\rangle \langle\psiup|=P_n
(\lambda)|\Phi_n\rangle \langle\Phi_n|$, where, $P_n
(\lambda)=(n+1)\lambda^n/(1+\lambda )^{n+2}$, $\lambda=\sinh^2\chi$,
which is half of the average number of photon pairs generated by one
pumping pulse and could be adjusted by the intensity of the pumping
pulsed. Therefore, PDCS is really just a photon-number states source
emitting n-photon pairs $|\Phi_n\rangle$ with probability $P_n
(\lambda)$. For implementations that do not apply phase
randomization, Eve may attack this QKD system more powerfully
\cite{phase randomization}, Therefore, for simplicity we assume that
Alice have applied phase randomization to her photon pairs.

  Here we focus on the attack to 2-photon pairs, because the 2-photon
pairs are dominant among the multi-photon pairs. For the practical
implementation \cite{DFS experiment} by Pan, Alice delays b mode of
the two spatial outputs of PDCS with $\Delta t_p$. Then through
phase-modulation by Pockel cells \cite{DFS experiment} 2-photon
pairs states could be described in creation operators form like
this:
\begin{equation}
\begin{aligned}
|-\rangle=\frac{1}{2\sqrt{3}}({H_a^{+}}^2 {V_b^{+}}^2-2H_a^{+}V_a^{+}H_b^{+}V_b^{+}+{V_a^{+}}^2 {H_b^{+}}^2)|vacuum\rangle\\
|+\rangle=\frac{1}{2\sqrt{3}}({H_a^{+}}^2 {V_b^{+}}^2+2H_a^{+}V_a^{+}H_b^{+}V_b^{+}+{V_a^{+}}^2 {H_b^{+}}^2)|vacuum\rangle\\
|0\rangle=\frac{1}{2\sqrt{3}}({H_a^{+}}^2 {V_b^{+}}^2+2iH_a^{+}V_a^{+}H_b^{+}V_b^{+}-{V_a^{+}}^2 {H_b^{+}}^2)|vacuum\rangle\\
|1\rangle=\frac{1}{2\sqrt{3}}({H_a^{+}}^2 {V_b^{+}}^2-2iH_a^{+}V_a^{+}H_b^{+}V_b^{+}-{V_a^{+}}^2 {H_b^{+}}^2)|vacuum\rangle\\
\end{aligned}
\end{equation}
where, $H_a^{+}$, $H_b^{+}$, $V_a^{+}$, and $V_b^{+}$ represent the
creation operators for horizontal polarized photons in $a$ mode,
horizontal polarized photons in $b$ mode, vertical polarized photons
in $a$ mode and vertical polarized photons in $b$ mode. For
simplicity, we assume Eve add a beam splitter (BS) to the both modes
$a$ and $b$ and we name the two spatial mode of the output of the BS
is $1$ and $2$. Now Eve has 4 spatial-temporal modes $a1$, $a2$,
$b1$ and $b2$, and creator operators for horizontal-polarized and
vertical-polarized photons in these new modes are correlated to
modes $a$ and $b$ by $H_a^{+}=(1/\sqrt{2})(H_{a1}^{+}-H_{a2}^{+})$,
$V_a^{+}=(1/\sqrt{2})(V_{a1}^{+}-V_{a2}^{+})$,
$H_b^{+}=(1/\sqrt{2})(H_{b1}^{+}-H_{b2}^{+})$, and
$V_a^{+}=(1/\sqrt{2})(V_{b1}^{+}-V_{b2}^{+})$. Then Eve can
post-select the states that each of modes $a1$, $b1$ $a2$ and $b2$
has one and only one photon respectively. We should notice that:
although through just one BS the probability of success of this
post-selection is just 1/4, Eve may use many BSs to make sure that
this probability will be close to 1. And states $|-\rangle$,
$|+\rangle$, $|0\rangle$ and $|1\rangle$ will be transformed to:

\begin{equation}
\begin{aligned}
|-\rangle'=\frac{1}{2\sqrt{2}}((H_{a1} V_{b1}-V_{a1}H_{b1})(H_{a2}
V_{b2}-V_{a2}H_{b2})\\+(H_{a1}V_{b2}-V_{a1}H_{b2})(H_{a2}V_{b1}-V_{a2}H_{b1}))\\
|+\rangle'=\frac{1}{2\sqrt{2}}((H_{a1} V_{b1}+V_{a1}H_{b1})(H_{a2}
V_{b2}+V_{a2}H_{b2})\\+(H_{a1}V_{b2}+V_{a1}H_{b2})(H_{a2}V_{b1}+V_{a2}H_{b1}))\\
|0\rangle'=\frac{1}{2\sqrt{2}}((H_{a1} V_{b1}+iV_{a1}H_{b1})(H_{a2}
V_{b2}+iV_{a2}H_{b2})\\+(H_{a1}V_{b2}+iV_{a1}H_{b2})(H_{a2}V_{b1}+iV_{a2}H_{b1}))\\
|1\rangle'=\frac{1}{2\sqrt{2}}((H_{a1} V_{b1}-iV_{a1}H_{b1})(H_{a2}
V_{b2}-iV_{a2}H_{b2})\\+(H_{a1}V_{b2}-iV_{a1}H_{b2})(H_{a2}V_{b1}-iV_{a2}H_{b1}))
\end{aligned}
\end{equation}
where $H(V)_{X}$ represents state vector $|H(V)\rangle_X$ for
abbreviation and the same below. Then Eve could use a unitary
transformation $\mathscr{U}_1$ to the photons in modes $a1$ and
$b2$. The definition of $\mathscr{U}_1$ is given by
$\mathscr{U}_1HVE_0=HVE_1$, $\mathscr{U}_1VHE_0=VHE_1$,
$\mathscr{U}_1HHE_0=HHE_2$ and $\mathscr{U}_1VVE_0=VVE_2$, in which
$E_x$ is an assist state of Eve and satisfying $\langle
E_0|E_1\rangle=\langle E_0|E_2\rangle=\langle E_1|E_2\rangle=0$. Eve
post-select $E_1$ through projection $P_1=|E_1\rangle\langle E_1|$
and then the four states will be mapped into the below states with
probability $75\%$.

\begin{equation}
\begin{aligned}
|-\rangle''=\frac{1}{\sqrt{5}}(2|X\rangle-|Y\rangle)\\
|+\rangle''=\frac{1}{\sqrt{5}}(2|X\rangle+|Y\rangle)\\
|0\rangle''=\frac{1}{\sqrt{5}}(2|X'\rangle+i|Y\rangle)\\
|1\rangle''=\frac{1}{\sqrt{5}}(2|X'\rangle+i|Y\rangle)
\end{aligned}
\end{equation}
where,
$|X\rangle=(1/\sqrt{2})(H_{a1}V_{b1}H_{a2}V_{b2}+V_{a1}H_{b1}V_{a2}H_{b2})$,
$|Y\rangle=(1/\sqrt{2})(H_{a1}H_{b1}V_{a2}V_{b2}+V_{a1}V_{b1}H_{a2}H_{b2})$
and
$|X'\rangle=(1/\sqrt{2})(H_{a1}V_{b1}H_{a2}V_{b2}-V_{a1}H_{b1}V_{a2}H_{b2})$.
Now, Eve can construct another unitary transformation
$\mathscr{U}_2$ defined by:
$\mathscr{U}_2|X\rangle|E_0\rangle=(\sqrt{3}|Z\rangle|E_1\rangle+|X\rangle|E_2\rangle)/2$,
$\mathscr{U}_2|X'\rangle|E_0\rangle=(\sqrt{3}|Z\rangle|E_3\rangle+|X'\rangle|E_2\rangle)/2$
and $\mathscr{U}_2|Y\rangle|E_0\rangle=|Y\rangle|E_2\rangle$. Here,
$|E\rangle$ represents an assist states of Eve and $\langle
E_0|E_1\rangle=\langle E_0|E_2\rangle=\langle E_1|E_2\rangle=0$. And
$|Z\rangle$ is any states of photon in modes $a1$, $b1$, $a2$ and
$b2$. With $\mathscr{U}_2$ and projection operation
$P_2=|E_2\rangle\langle E_2|$, the four photon states will be mapped
to the followed form with probability 40\%.

\begin{equation}
\begin{aligned}
|-\rangle'''&=\frac{1}{\sqrt{2}}(|X\rangle-|Y\rangle)\\
&=\frac{1}{\sqrt{2}}(H_{a1}V_{b2}-V_{a1}H_{b2})\frac{1}{\sqrt{2}}(H_{a2}V_{b1}-V_{a2}H_{b1})\\
|+\rangle'''&=\frac{1}{\sqrt{2}}(|X\rangle+|Y\rangle)\\
&=\frac{1}{\sqrt{2}}(H_{a1}V_{b2}+V_{a1}H_{b2})\frac{1}{\sqrt{2}}(H_{a2}V_{b1}+V_{a2}H_{b1})\\
|0\rangle'''&=\frac{1}{\sqrt{2}}(|X\rangle+i|Y\rangle)\\
&=\frac{1}{\sqrt{2}}(H_{a1}V_{b2}+iV_{a1}H_{b2})\frac{1}{\sqrt{2}}(H_{a2}V_{b1}+iV_{a2}H_{b1})\\
|1\rangle'''&=\frac{1}{\sqrt{2}}(|X\rangle-i|Y\rangle)\\
&=\frac{1}{\sqrt{2}}(H_{a1}V_{b2}-iV_{a1}H_{b2})\frac{1}{\sqrt{2}}(H_{a2}V_{b1}-iV_{a2}H_{b1})\\
\end{aligned}
\end{equation}
 Obviously, with the states $|-\rangle'''$, $|+\rangle'''$, $|0\rangle'''$ and
$|1\rangle'''$, Eve can keep one pair and send the other pair to Bob
through a special channel controlled by herself. When Alice and Bob
do basis reconciliation, Eve will get all secret information. This
is just the same as PNS attack \cite{PNS1,PNS2,PNS3}.

 Let us review our attack strategy. First, Eve divides the two
photons in modes $a$ and $b$ into modes $a1$, $a2$ and $b1$, $b2$
respectively. With many BSs, success probability of this step is
close to 1. Second, Eve applys unitary transformation
$\mathscr{U}_1$ and projection $P_1$, she gets an intermediate state
with success probability $75\%$. Finally, She applys unitary
transformation $\mathscr{U}_2$ and projection $P_2$, she gets the
final state which she can launch PNS attack immediately and success
probability of this step is $40\%$. Overall, for 2-photon pairs Eve
will launch PNS attack with probability of $75\%\times 40\%=30\%$ or
discard a failure case with probability $1-30\%=70\%$.

 According to the above fact and the discussion of Ref. \cite{PNS1, PNS2, PNS3}, we
know the security distance ($L$) of this scheme must obey
$P_1(\lambda)(10^{-kL/10})^2\geq P_2(\lambda)\times 30\%$ in which
$k$ is the transmission fiber loss constance. If we assume
$k=0.2dB/km$ which is a typical value of this constance and
$\lambda=0.1$, we obtain $L\leqq 37.4 km$. This is a highly
unsatisfactory situation. How to prolong the  security distance is
what we will discuss in the next section.

\section{Decoy states to Boileau's scheme}

The rate of secret key bits ($R$) for BB84 protocol with nonideal
source can be determined by GLLP \cite{GLLP}:
\begin{equation}
\begin{aligned}
R\geqslant R^L=q[-Q_\lambda
f(E_\lambda)H_2(E_\lambda)+P_1(\lambda)S_1^L(1-H_2(e_1^U))]
\end{aligned}
\end{equation}
Here, $R^L$ represents the lower bound of $R$, q depends on protocol
($1/2$ for Boileau's scheme), $Q_\lambda$ is the overall counting
rate for the photon pairs, $\lambda$ is half of the average number
of the photon pairs, $f(E_\lambda)$ is error correction efficiency,
$E_\lambda$ is the quantum bit error rate (QBER) of the key bit,
$H_2$ is the binary Shannon information function, $S_1$ is the
counting rate for the 1-photon pairs, and $e_1$ is the QBER of the
key bits generated by the 1-photon pairs. Similar to BB84 based on
weak coherent states, we need to modulate $\lambda$ to several
values randomly. Through watching counting rates for different
$\lambda$, we can obtain the lower bound of $S_1$ ($S_1^L$) and the
upper bound of $e_1$ ($e_1^U$). Finally, $R^L$ can be obtained by
equation (7).

 Our 3-intensity protocol is: Alice randomly emits photon pairs of
density matrix $\rho_\lambda$, $\rho_{\lambda'}$, and $0$ ($\lambda$
for signal states , $\lambda'$ ($\lambda>\lambda'$) and $0$ for
decoy states) , then Bob can get their counting rates $Q_\lambda$,
$Q_{\lambda'}$ and $S_0$. With formulas we derived later, $S_1^L$
and $e_1^U$ can be obtained. Finally, $R^L$ is given by equation
(7). Now we drive these formulas.

 The counting rates for the two intensity ($\lambda$ and $\lambda'$) photon pairs is determined by:
\begin{equation}
\begin{aligned}
Q_\lambda=\sum_{n=0}^{\infty}P_n(\lambda)S_n
\end{aligned}
\end{equation}

\begin{equation}
\begin{aligned}
Q_{\lambda'}=\sum_{n=0}^{\infty}P_n(\lambda')S_n
\end{aligned}
\end{equation}
where, $S_n$ represents the counting rate for n-photon pair states
$|\Phi_n\rangle$. Then QBER for the $\lambda$ ($E_{\lambda}$) is
determined by:
\begin{equation}
\begin{aligned}
E_{\lambda}Q_{\lambda}=\sum_{n=0}^{\infty}e_nP_n(\lambda)S_n
\end{aligned}
\end{equation}
In which, $e_n$ is the QBER of the key bits generated by the
n-photon pairs $|\Phi_n\rangle$. Before the derivation of the
formula to calculate $S_1^L$ and $e_1^U$, we prove that
$\frac{P_2(\lambda)}{P_2(\lambda')}P_n(\lambda')\leqslant
P_n(\lambda)$ for all of $n\geqslant 2$.

\begin{equation}
\begin{aligned}
&\frac{P_2(\lambda)}{P_n{(\lambda)}}-\frac{P_2(\lambda')}{P_n{(\lambda'})}\\
&=\frac{3}{n+1}((1+\frac{1}{\lambda})^{n-2}-(1+\frac{1}{\lambda'})^{n-2})
&\leqslant 0
\end{aligned}
\end{equation}
With this result, we can deduce the formula for calculating $S_1^L$:
\begin{equation}
\begin{aligned}
Q_\lambda&=P_0(\lambda)S_0+P_1(\lambda)S_1+P_2(\lambda)S_2+P_3(\lambda)S_3+\cdots\\
&\geqslant
P_0(\lambda)S_0+P_1(\lambda)S_1+\frac{P_2(\lambda)}{P_2(\lambda')}\sum_{n=2}^{\infty}P_n(\lambda')S_n
\end{aligned}
\end{equation}
With equation (9), we have:
\begin{equation}
\begin{aligned}
S_1^L=\frac{(P_2(\lambda')P_0(\lambda)-P_2(\lambda)P_0(\lambda'))S_0+P_2(\lambda)Q_{\lambda'}-P_2(\lambda')Q_{\lambda}}{P_2(\lambda)P_1(\lambda')-P_2(\lambda')P_1(\lambda)}
\end{aligned}
\end{equation}
According to equation (10) and \cite{decoy theory4}, $e_1^U$ can be
given by:
\begin{equation}
\begin{aligned}
e_1^U=\frac{(E_{\lambda}Q_{\lambda}-\frac{S_0P_0(\lambda)}{2})}{P_1(\lambda)S_1^L}
\end{aligned}
\end{equation}

 With equation (13) and (14), $S_1^L$ and
$e_1^U$ can be obtained. Finally, $R^L$ is given by equation (7).

 For experiment, 2-intensity decoy states protocol is quite
convenient \cite{decoy experiment7}. In this case, Alice randomly
emits photon pairs of density matrix $\rho_\lambda$ for signal
states, $\rho_{\lambda'}$ for decoy states, then Bob can get their
counting rates $Q_\lambda$, $Q_{\lambda'}$. We now deduce the
formula to calculate $S_1^L$ and $e_1^U$ just from $Q_\lambda$,
$Q_{\lambda'}$.

 According to equation (10), the upper bound of $S_0$ ($S_0^U$) can
be given by:
\begin{equation}
\begin{aligned}
S_0^U=\frac{2E_{\lambda}Q_{\lambda}}{P_0(\lambda)}
\end{aligned}
\end{equation}
Then from equation (10), $S_1^L$ for two-intensity case can be given
by:

\begin{equation}
\begin{aligned}
&S_1^L\\
&=\frac{2(P_2(\lambda')P_0(\lambda)-P_2(\lambda)P_0(\lambda'))\frac{E_{\lambda}Q_{\lambda}}{P_0(\lambda)}+P_2(\lambda)Q_{\lambda'}-P_2(\lambda')Q_\lambda}{(P_2(\lambda)P_1(\lambda')-P_2(\lambda')P_1(\lambda))P_0(\lambda)}
\end{aligned}
\end{equation}
To get $e_1^U$ for two intensity case, we just set lower bound of
$S_0$ ($S_0^L$) to be $0$, then with equation (10) and (16), $e_1^U$
is given by:

\begin{equation}
\begin{aligned}
e_1^U=\frac{E_{\lambda}Q_{\lambda}}{P_1(\lambda)S_1^L}
\end{aligned}
\end{equation}

 Equations (16) and (17) are for 2-intensity case. With these
equations, we have established the basic methods to beat PNS attack
in Boileau's QKD scheme. Next, we will make sure that this decoy
states method can improve the performance of Boileau's QKD scheme
impressively.

\section{improvement by decoy states}

 Now, we will show the improvement for the performance by the
introduction of decoy states through the numerical simulations. In
the followed discussions and simulations, we neglect the error
induced by channel and assume Bob's measurements are perfect except
a few dark counts for simplicity. According to Ref. \cite{DFS
experiment}, Bob's measurement is equivalent to the projection to
the polarization states $F$ and $S$ defined by $H=(F+S)/\sqrt{2}$
and $V=(F-S)/\sqrt{2}$ respectively. We rewrite the encoding states
$|+\rangle$ and $|-\rangle$ in the form of $F$ and $S$:
$|+\rangle=\frac{1}{\sqrt{n+1}}\sum_{m=0}^n(-1)^mF_a^{n-m}S_a^mF_b^{n-m}S_a^m$,
$|-\rangle=\frac{1}{\sqrt{n+1}}\sum_{m=0}^n(-1)^mF_a^{n-m}S_a^mF_b^mS_a^{n-m}$.
For Bob, if he observers the $F_aS_b$ or $S_aF_b$, it's will be
$|+\rangle$ while the $F_aF_b$ or $S_aS_b$ is for the result of
$|+\rangle$. According to Ref.\cite{decoy theory4}, the transmission
efficiency for the n-photon pulses $\eta_n$ can be written as
$\eta_n=1-(1-\eta)^n$, in which $\eta$ is the transmission
efficiency of the fiber channel and $\eta=10^{(-kL/10)}$, $K$ is the
transmission fiber loss constance and L is the fiber length. Since
our goal is to show the difference between the original Boileau's
scheme and this scheme with decoy states but not the exact $R^L$
verse fiber length, we take the efficiency of the detector and loss
due to projection to the DFS space or other causes just as a part of
fiber loss and don't care these values. We assume the dark counting
rates of the detectors is $D$. Since Bob must neglect all the three
or four folds counts, $S_n$ can be written as:
\begin{equation}
\begin{aligned}
S_n&=\frac{(1-D)^2}{n+1}\sum_{m=0}^n((\eta_{n-m}(1-\eta)^m+\eta_m(1-\eta)^{n-m})^2\\
&+4\eta_{n-m}(1-\eta)^m(1-\eta)^nD+4\eta_{m}(1-\eta)^{n-m}(1-\eta)^nD\\
&+4(1-\eta)^{2n} D^2)
\end{aligned}
\end{equation}
 Then with equation (8), we can get the formulas to estimate the
$Q_\lambda$ and $Q_{\lambda'}$.

\begin{equation}
\begin{aligned}
Q_{\lambda}&=\sum_{n=0}^{\infty}P_n(\lambda)S_n\\&=\frac{2(1-D)^2}{(1+\lambda\eta(3-\eta)+\lambda^2\eta^2(2-\eta))^2}\times
\\
&(4\lambda\eta
D(1-\eta)(1+\lambda\eta)+2D^2(1+\lambda\eta)^2\\
&+\lambda\eta^2(1+\lambda^2(2-\eta)\eta+\lambda(\eta^2-2\eta+3)))
\end{aligned}
\end{equation}

 For simplicity we neglect the probability that a survived photon
hitting a wrong detector, then $e_n$ is written like:
\begin{equation}
\begin{aligned}
e_nS_n&=[\sum_{m=0}^n(2\eta_{n-m}(1-\eta)^m(1-\eta)^{n-m}\eta_m+2\eta_m(1-\eta)^{n-m}(1-\eta)^nD\\
&+2\eta_{n-m}(1-\eta)^m(1-\eta)^nD+(1-\eta)^{2n}2D]\frac{(1-D)^2}{n+1}
\end{aligned}
\end{equation}
in which, the first term of the summation corresponds to the case of
the photons in modes $a$ and $b$ both hitting the detectors. Only
when $n\geqslant 2$, this term does not equal to 0. The second and
third terms in above summation represent to the case photons in only
one mode ($a$ or $b$) hit the detector. The dark count of one
detector may result in QBER in this situation. The last term of the
summation is for the case of all the photons are absorded by fiber.

 With this, we can estimate the QBER $E_\lambda$ as:
\begin{equation}
\begin{aligned}
E_\lambda&=\sum_{n=0}^{\infty}P_n(\lambda)e_nS_n/Q_\lambda\\
&=(D+\lambda D\eta+\lambda\eta(1-\eta))^2\times\\
&[4\lambda D\eta(1-\eta)(1+\lambda\eta)+2D^2(1+\lambda\eta)^2\\
&+\lambda\eta^2(1+\lambda^2\eta(2-\eta)+\lambda(\eta^2-2\eta+3))]^{-1}
\end{aligned}
\end{equation}
 Now with equations (19) and (21) and setting $k=0.2 dB/km$,
$D=10^{-6}/pulse$, and $f(E_\lambda)=1.2$, the $Q_\lambda$,
$Q_{\lambda'}$, $E_\lambda$ can be calculated by numerical
simulations. Then with equations (13) and (14), the $S_1^L$ and
$e_1^U$ can be obtained. Finally, the relation between $R^L$ and
fiber length $L$ can be get. And the results are depicted in Fig. 1.
In Fig. 1, the solid curve is for the case that no decoy states is
employed. In this case, for the calculation of $S_1^L$ and $e_1^U$
we have to assume that $S_n=1 (n\geqslant 2)$ and with equation
(15), then obviously the $S_1$ is given by:
\begin{equation}
\begin{aligned}
S_1^L&=\frac{Q_\lambda-P_0(\lambda)S_0^U-\sum_{n=2}^\infty P_2(\lambda)}{P_1(\lambda)}\\
&=\frac{Q_\lambda(1-2E_\lambda)-(1-P_0(\lambda)-P_1(\lambda))}{P_1(\lambda)}
\end{aligned}
\end{equation}
\begin{figure}[!h]\center
\resizebox{8.5cm}{!}{\includegraphics{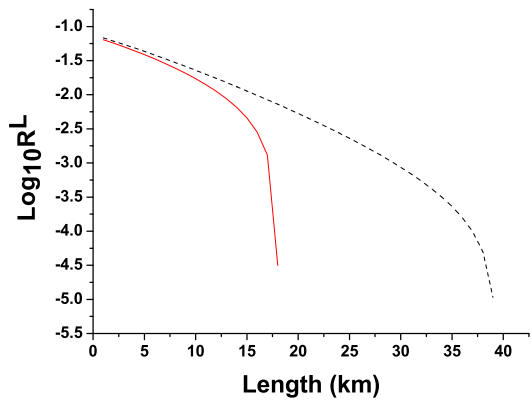}} \caption{Lower
bound of secret bit generation rate ($R_1^L$) verse fiber length
$L$. Solid curve: no decoy states is employed, Alice just emits PDCS
with half average number of photon pair $\lambda=0.1$. Dashed curve:
for the 3-intensity case, Alice randomly used PDCS with half average
number of photon pair $\lambda=0.1$, $\lambda'=0.01$ and
0.}\label{schematic}
\end{figure}
The $e_1^U$ is then calculated by equation (14). With this method,
$R^L$ is obtained by equation (7). From Fig. 1, we found that the
3-intensity decoy states method can improve the performance of
Boileau's scheme dramatically. The longest security distance in
original Boileau's scheme is about 18 km while this distance for
3-intensity decoy states method will be 40 km. This improvement
means about the 4.4dB increase in longest security distance.

\section{Conclusion}

 According to above discussions, we proved that through the
introduction of decoy states method, especially the 3-intensity
decoy states, the performance of Boileau's DFS type QKD would be
dramatically improved. Thanks to 3-intensity decoy state protocol
the increase of longest security distance can be 4.4dB. This
increase relays on the ability of 3-intensity decoy states protocol
can obtain a tighter bound of  $S_1^L$ and $e_1^U$. Furthermore one
can estimate the information leaked to Eve with high precision and
higher key bit rate and longer security distance can be obtained. We
hope that our protocol could be implemented soon.

 This work was supported by National Fundamental
Research Program of China (2006CB921900), National Natural Science
Foundation of China (60537020, 60621064) and the Innovation Funds of
Chinese Academy of Sciences. To whom correspondence should be
addressed, Email: zwzhou@ustc.edu.cn and zfhan@ustc.edu.cn.


\begin{thebibliography}{00}


\bibitem{BB84}
C. H. Bennett,  G.Brassard, Proceedings of \emph{IEEE International
Conference on Computers, Systems, and Signal Processing}, (IEEE,
1984), pp. 175-179.
\bibitem{ekert1991}
A. K. Ekert, Phys. Rev. Lett. \textbf{67}, 661 (1991)
\bibitem{Gisin}
N. Gisin et al., Rev. Mod. Phys. \textbf{74}, 145 (2002)
\bibitem{Walton}
Z. D. Walton, A. F. Abouraddy, A.V. Sergienko, B. E. A. Saleh, and
M.C. Teich, Phys. Rev. Lett. \textbf{91}, 087901 (2003)
\bibitem{Boileau}
J.-C. Boileau, R. Laflamme, M. Laforest, and C. R. Myers, Phys. Rev.
Lett. \textbf{93}, 220501 (2004)
\bibitem{DFS experiment}
Teng-Yun Chen, Jun Zhang, J.-C. Boileau, Xian-Min Jin, Bin Yang,
Qiang Zhang, Tao Yang, R. Laflamme, and Jian-Wei Pan, Phys. Rev.
Lett. \textbf{96}, 150504 (2006)
\bibitem{qkd1}
M. Bourennane et al., Opt. Express \textbf{4}, 383 (1999)
\bibitem{qkd2}
D. Stucki et al., New. J. Physics, \textbf{4}, 41, (2002)
\bibitem{qkd3}
H. Kosaka et al., Electron. Lett. \textbf{39}, 1199 (2003)
\bibitem{qkd4}
C. Gobby, Z.L. Yuan, and A.J. Shields, Appl. Phys. Lett.
\textbf{84}, 3762 (2004);
\bibitem{F-M}
X.-F. Mo et al., Optics Letters, Vol. \textbf{30}, Issue 19, pp.
2632-2634 (October 2005)
\bibitem{PNS1}
B. Huttner, N. Imoto, N. Gisin, and T. Mor, Phys. Rev. A
\textbf{51}, 1863 (1995);
\bibitem{PNS2}
G. Brassard et al., Phys. Rev. Lett. \textbf{85}, 1330 (2000).
\bibitem{PNS3}
N. Lu¡§tkenhaus, Phys. Rev. A \textbf{61}, 052304 (2000).
\bibitem{decoy theory1}
W.-Y. Hwang, Phys. Rev. Lett. \textbf{91}, 057901 (2003).
\bibitem{decoy theory2}
H.-K. Lo, X. Ma, and K. Chen, Phys. Rev. Lett. \textbf{94}, 230504
(2005).
\bibitem{decoy theory3}
X.-B. Wang, Phys. Rev. Lett. \textbf{94}, 230503 (2005);
\bibitem{decoy theory4}
X. Ma et al., Phys. Rev. A \textbf{72}, 012326 (2005).
\bibitem{decoy experiment1}
Y. Zhao et al., Phys. Rev. Lett. \textbf{96}, 070502 (2006)
\bibitem{decoy experiment2}
Yi Zhao et al, Proceedings of IEEE International Symposium on
Information Theory 2006, pp. 2094-2098
\bibitem{decoy experiment3}
C.-Z. Peng et al., Phys. Rev. Lett. \textbf{98}, 010505 (2007)
\bibitem{decoy experiment4}
D. Rosenberg, J. W. Harrington, P. R. Rice, et al., Phys. Rev. Lett.
\textbf{98}, 010503 (2007)
\bibitem{decoy experiment5}
Z. L. Yuan, A. W. Sharpe, and A. J. Shields, Appl. Phys. Lett.
\textbf{90} 011118 (2007)
\bibitem{decoy experiment6}
Tobias Schmitt-Manderbach et al., Phys. Rev. Lett. \textbf{98},
010504 (2007)
\bibitem{decoy experiment7}
Z.-Q. Yin et al., quant-ph/0704.2941 (2007)
\bibitem{decoy theory5}
X.-B. Wang, Phys. Rev. A \textbf{72}, 012322 (2005)
\bibitem{nonorthogonal state protocol}
J.-B. Li, and X.-M. Fang, Chin. Phys. Lett. \textbf{23}, No. 4
(2006)
\bibitem{photon-number-resolving method}
Qing-yu Cai, and Yong-gang Tan, Phys. Rev. A. \textbf{73}, 032305
(2006)
\bibitem{herald1}
Tomoyuki Horikiri, and Takayoshi Kobayashi, Phys. Rev. A
\textbf{73}, 032331 (2006)
\bibitem{herald2}
Qin Wang, X.-B. Wang, and G.-C. Guo, Phys. Rev. A \textbf{75},
012312 (2007)
\bibitem{MCS}
Z.-Q. Yin, Z.-F. Han, F.-W. Sun, and G.-C. Guo, Phys. Rev. A
\textbf{76}, 014304 (2007)
\bibitem{PDCS}
Xiongfeng Ma, Chi-Hang Fred Fung, and Hoi-Kwong Lo, Phys. Rev. A
\textbf{76}, 012307 (2007)
\bibitem{GLLP}
D. Gottesman, H.-K. Lo, N. L\"{u}tkenhaus, and J. Preskill, Quantum
Inf. Comput. \textbf{4}, 325 (2004).
\bibitem{intensity error1}
X.-B. Wang, C.-Z. Peng, and J.-W. Pan, Appl. Phys. Lett.
\textbf{90}, 031110 (2007)
\bibitem{intensity error2}
X.-B. Wang, Phys. Rev. A \textbf{75}, 052301 (2007)
\bibitem{phase randomization}
Hoi-Kwong Lo and John Preskill, arXiv:quant-ph/0504209 (2005)
\end{thebibliography}
\end{document}